# Incommensurate atomic and magnetic modulations in the spin-frustrated β-NaMnO$_2$ triangular lattice.

Fabio Orlandi,[1] Eleni Aza,[2,3] Ioanna Bakaimi,[2,†] Klaus Kiefer,[4] Bastian Klemke,[4] Andrej Zorko,[5] Denis Arčon,[5,6] Christopher Stock,[7] George D. Tsibidis,[2] Mark A. Green,[8] Pascal Manuel,[1] and Alexandros Lappas[2,*]

[1] ISIS Facility, Rutherford Appleton Laboratory, Harwell Oxford, Didcot OX11 0QX, United Kingdom

[2] Institute of Electronic Structure and Laser, Foundation for Research and Technology–Hellas, Vassilika Vouton, 71110 Heraklion, Greece

[3] Department of Materials Science and Engineering, University of Ioannina, 451 10 Ioannina, Greece

[4] Department Sample Environment and CoreLab Quantum Materials, Helmholtz-Zentrum Berlin für Materialien und Energie GmbH, D-14109 Berlin, Germany

[5] Jozef Stefan Institute, Jamova c. 39, 1000 Ljubljana, Slovenia

[6] Faculty of Mathematics and Physics, University of Ljubljana, Jadranska c. 19, 1000 Ljubljana, Slovenia

[7] School of Physics and Astronomy, University of Edinburgh, Edinburgh EH9 3JZ, United Kingdom

[8] School of Physical Sciences, University of Kent Canterbury, Kent CT2 7NH, United Kingdom

**ABSTRACT.** The layered β-NaMnO$_2$, a promising Na-ion energy-storage material has been investigated for its triangular lattice capability to promote complex magnetic configurations that may release symmetry restrictions for the coexistence of ferroelectric and magnetic orders. The complexity of the neutron powder diffraction patterns underlines that the routinely adopted commensurate structural models are inadequate. Instead, a single-phase superspace symmetry description is necessary, demonstrating that the material crystallizes in a compositionally modulated **q**= (0.077(1), 0, 0) structure. Here, Mn$^{3+}$ Jahn-Teller distorted MnO$_6$ octahedra form corrugated layer stacking sequences of the β-NaMnO$_2$ type, which are interrupted by flat sheets of the α-like oxygen topology. Spontaneous long-range collinear antiferromagnetic order, defined by the propagation vector **k**= (½, ½, ½), appears below $T_{N1}$= 200 K. Moreover, a second transition into a spatially modulated proper-screw magnetic state (**k**±**q**) is established at $T_{N2}$= 95 K, with an antiferromagnetic order parameter resembling that of a two-dimensional (2D) system. The evolution of $^{23}$Na NMR spin-lattice relaxation identifies a magnetically inhomogeneous state in the intermediate $T$-region ($T_{N2}$ <$T$< $T_{N1}$), while its strong suppression below $T_{N2}$ indicates that a spin-gap opens in the excitation spectrum. High-resolution neutron inelastic scattering confirms that the magnetic dynamics are indeed gapped (Δ~5 meV) in the low-temperature magnetic phase, while simulations on the basis of the single-mode approximation suggest that Mn-spins residing on adjacent antiferromagnetic chains, establish sizable 2D correlations. Our analysis points that novel structural degrees of freedom promote, cooperative magnetism and emerging dielectric properties in this non-perovskite-type of manganite.

## I. INTRODUCTION.

Devising cost-efficient chemical routes for multiferroic magnetoelectric compounds that foster coupling between spins and other electron degrees of freedom is a fascinating problem of both fundamental and technological interest.[1] Engineering the materials' structure to accommodate unusual coordinations of interacting neighbours offers one such viable, but challenging avenue. The perturbation of exchange interactions that emerge from competition due to magnetic frustration, [2, 3, 4] can select complex spin arrangements that release symmetry restrictions and realize the long-wanted coupling of otherwise mutually exclusive ferroelectric and magnetic orders. In this context, the non-perovskite, two-dimensional (2D) Na-Mn-O oxides are investigated as a testing ground for such a kind of magnetoelectricity. These are rock-salt derivatives of the family $A^+Me^{3+}O_2$ (A= alkali metal, Me= 3d transition metal) delafossites [5, 6] that have attracted considerable interest due to their physical and chemical properties. They include, transparent conducting oxides, such as the $CuAlO_2$ [7], superconductors, like the hydrated variant $Na_{0.3}CoO_2 \cdot 1.3H_2O$ [8] of the P2-$Na_yCoO_2$ bronzes [9], multiferroics as $AFeO_2$ (A= Na, Ag)[10, 11], and cathodic materials for high-capacity Na-ion re-chargeable batteries, like P2-$Na_yMn_{1-x}M_xO_2$ (x, y ≤ 1, M = Ni, Mg, Li) [12]. Such intercalation materials show high structural flexibility upon alkali metal insertion or extraction and give rise to a rich phase diagram. [13] The crystal chemistry of $AMeO_2$ allows for polymorphism due to oxygen-layer gliding processes.[14] Consequently, their performance is mediated by phase transitions between nearly degenerate structural types (e.g. designated, as O3- (3R; R-3m) and P2- (P6$_3$/mmc)), [12,15] while extended defects (e.g. stacking faults) formed between various crystal domains, render the apparently simple $A_xMeO_2$ bronzes metastable. Therefore, new insights on the impact of their inherent compositional variation are sought in order to explain their complicated sequences of electronic and structural processes.

Core concepts of materials science point out that when near-degenerate energy states are involved, compositional modulation[16] often emerges as a naturally evolving process that relieves frustration by satisfying the cation-anion chemical requirements, as for example in ferroelectrics, [17] and shape memory alloys. [18, 19] Then, alternatives to traditional crystallographic approaches are necessary in order to understand how subtle structural modulations in correlated transition metal oxides (e.g. cation order and tilting of metal-oxygen coordination polyhedral etc.) entangle their electron degrees of freedom and lead to novel behaviour, extending from heterogeneous catalysis and spin-induced ferroelectricity to high-temperature superconductivity. The ability to control such functional properties, often emerging in the framework of broken symmetries (as in $TbMnO_3$ and $Ni_3V_2O_8$ magnetoelectric materials), [20] relies in understanding the role of residual disorder governing the modulation of atomic positions and magnetic moments. The superspace formalism, previously implemented for the description of modulated chemical crystal structures,[21] has grown as a powerful method especially when nuclear and magnetic modulations intertwine in the same phase. [22] Diverse structural types, ranging from perovskites ($CaMn_7O_{12}$ [23], $Pb_2MnWO_6$ [24]) to wolframite-type ($MnWO_4$ [25]) modulated structures, which all display symmetry-allowed coupling of electric polarization and magnetization, are illustrative examples of the importance of a robust and efficient treatment of the symmetry of nuclear and magnetic modulations.

The focus here is on two particular polymorphs in the Na-Mn-O system which crystallizes in distorted variants of the O3-$NaFeO_2$ structure (3R polytype, *R-3m*). [26] In these layered compounds the spontaneous deformation of the $MnO_6$ octahedra is caused by the Jahn-Teller effect, inherent to the high-spin $Mn^{3+}$ cations ($t_{2g}^3 e_g^1$; S= 2; $\mu_{eff} \cong 4.9$ $\mu_B$). Because of this distortion, α-$NaMnO_2$ becomes monoclinic (*C2/m*), with flat [27] $MnO_6$ sheets (Figure S1a) [28], whilst β-$NaMnO_2$ appears to adopt an orthorhombic cell (*Pmmn*), entailing zig-zag [29] $MnO_6$ sheets (Figure S1b) [28]. The latter polytype is similar to the thermodynamically stable lithiated analogue β-$LiMnO_2$, [30] an important precursor phase for cathode materials in solid-state Li-ion batteries. [31] Moreover, specific challenges facing the Mn-containing systems are governed, (a) by the very similar free-energies of the α- and β- $NaMnO_2$ polymorphs, [32] which suggest that intermediate phases with compositional modulations could be formed at a very low energy cost, and (b) by the Mn topology (see Figure S1, Supplemental Material) [28] that maps out a triangular lattice [33], inferring some degree of spin-frustration that renders these polymorphs sensitive to small perturbations.

In view of the former characteristic, transmission electron microscopy and synchrotron X-ray powder diffraction have shown that on the basis of superspace formalism, planar defects could act as a structure-directing mechanism in the cation-ordered rock-salt-type $AMeO_2$ structures, and in particular, the α- and β- phases of $NaMnO_2$ can be gradually transformed into each other by changing the density of the involved twin planes. [33] Interestingly, the presence of local intergrowths of β-polymorph and stacking faults within the lattice of the parent α-$NaMnO_2$ phase is shown to be controlled in single-crystals grown under optimal conditions.[34] This apparent energy degeneracy between α- and β- type oxygen coordinations seems to play an important role in determining the particularly high charge capacity (ca. 190 mA h g$^{-1}$) of polycrystalline β-$NaMnO_2$ as an earth-abundant Na-ion cathode. [35] As of the second inherent feature, neutron powder diffraction has shown that despite the considerable spin-frustration in α-$NaMnO_2$, Néel order sets-in at 45 K. [36] With this concomitant symmetry breaking, a spin-gap due to leading quasi-one dimensional interactions (with a predominant nearest-neighbor exchange interaction of $J_1$~ 72 K[37] and frustrated $J_2 \cong 0.44\ J_1$ [38]; Figure S1a [28]) describes the low-energy magnetic dynamics, while a peculiar magnetostructural inhomogeneity emerges as a consequence of the system's tendency to remove magnetic degeneracy due to spin frustration. [39, 40] On the other hand, the magnetic ground state of β-$NaMnO_2$ is less well understood from the experimental point of view. Theoretical calculations though, predict that a spin-model with two-dimensional couplings ($J_1$~ 70 K nearest neighbor and $J_3$~ 57 K next nearest neighbor; Figure S1b) [28] and a weaker frustrated interaction ($J_2$~ 13 K) are likely to describe the experimental magnetic susceptibility. This material also manifests an abundant quasi-periodic arrangement of defects. [33] Moreover, room-temperature $^{23}$Na solid-state nuclear magnetic resonance (NMR) spectra supported by first-principles DFT computations identified a wealth of local structural rearrangements, entailing a trade-off between the majority β-type nanodomains and those of the α-like phase upon electrochemical cycling of sodium.[41]

The present contribution provides a new powerful neutron powder diffraction insight on β-$NaMnO_2$, highlighting that this challenging material is stabilized by near-equivalent in energy lattice conformations. The strength of superspace formalism has been utilized to describe the structure on the basis of a single-phase model, entailing an incommensurate compositional modulation. The latter is depicted as a coherent intergrowth of two types of $NaMnO_2$ layers, reflecting the α- and β- type oxygen coordinations, and is shown to determine the material's physical properties. We illustrate the implications of the modified lattice topology, with its intrinsic extended defects, on the successive magnetic phase transitions. Furthermore, temperature-dependent $^{23}$Na NMR and inelastic neutron scattering experiments point that the magnetic dynamics are gapped, while the influence of the magnetic order on the electric dipole order is also reflected in the temperature- and field- dependent magnetocapacitance studies.

## II. EXPERIMENTAL METHODS.

Polycrystalline β-$NaMnO_2$ samples were synthesized by a high-temperature solid-state chemistry protocol reported before, [33] while phase identification was undertaken by x-Ray powder diffraction (XRPD) experiments carried out on a Rigaku D/MAX-2000H rotating Cu anode diffractometer. β-$NaMnO_2$ specimens were air-sensitive and all post-synthesis handling was carried with the aid of an Ar-circulating MBRAUN anaerobic glove box.

Dc magnetic susceptibility as a function of temperature (5 ≤ T ≤ 300 K) was measured on 20 mg batches of powder samples with a Superconducting Quantum Interference Device (SQUID) magnetometer (Quantum Design, MPMS-XL7) under a moderate magnetic field (H = 20 mT). Heat capacity ($C$) was measured at zero-field on a cold-pressed pelletized powder sample by means of the relaxation technique, utilizing a physical property measurement system (Quantum Design, PPMS).

NMR measurements on the $^{23}$Na nucleus (nuclear spin $I$= 3/2) were performed on a powder sample sealed in a pyrex sample holder. $^{23}$Na NMR spectra and spin-lattice relaxation rate $1/T_1$ were recorded between 50 and 300 K in a magnetic field of 8.9 T using a solid-echo and inversion recovery pulse sequences, respectively. Wide-line $^{23}$Na NMR powder data were obtained as sums of individual spectra acquired by changing the measurement frequency in 50 kHz steps over ±3 MHz around the $^{23}$Na reference frequency, $\nu_0$= 100.5234 MHz, which was determined from a 0.1 M NaCl solution. The spin-lattice relaxation rate measurements were performed at the position of the central line.

Neutron powder diffraction data were collected on the WISH diffractometer,[42] operating at the second target station (TS2) at the ISIS pulsed neutron source in the UK. WISH, with its high-brilliance, is particularly optimized for providing high resolution at long d-spacing required for magnetic studies. For this purpose, a 2.7 g polycrystalline sample was loaded in a 8 mm V-can, which was then sealed with indium wire inside a high-quality, He-circulating anaerobic glove box. An Oxford Instrument liquid helium cryostat was used for the temperature dependent diffraction experiments. Data analysis was performed by using the Jana2006 software[43] for the Rietveld refinements, whereas the group theory analysis was performed with the help of the ISODISTORT software.[44]

Inelastic neutron scattering work was performed on the MARI direct geometry chopper spectrometer (ISIS, UK) and also on the DCS spectrometer (NIST, USA). Experiments on MARI used incident energies $E_i$=85 and 150 meV, with a Gd Fermi chopper spun at 300 and 450 Hz, respectively. Measurements on DCS were done with an incident energy of $E_i$=14.2 meV. A 7.3 g of a powder sample was loaded in an annular aluminum sachet that was placed inside a cylindrical Al-can for the ISIS experiment, while a 5 g sample was loaded in V-can for the NIST experiment. In either case the cans were sealed with indium wire and they were cooled at low temperatures with a top-loaded closed-cycle refrigerator. All data has been corrected for background and also phonons from the structural lattice. For the MARI data, the background plus phonon contribution to the scattering at each energy transfer was estimated from the high angle detector banks where magnetic scattering is suppressed owing to the $Mn^{3+}$ form factor. We have fit the high angle and high momentum detectors at a fixed energy transfer to the form L(Q)= $L_0$ + $L_1 \times Q^2$, with $L_0$ capturing the background and $L_1$ providing an estimate of the phonon scattering. L(Q) was then used to estimate the background and phonon scattering at low momentum transfers and then it was subtracted. For data taken on DCS, the background was estimated by using the requirement for detailed balance as discussed previously.[45]

The dielectric permittivity of ~3 mm pellets of pressed polycrystalline samples, without electrodes attached on the two flat surfaces, was studied at the CoreLab for Quantum Materials in the Helmholtz-Zentrum, Berlin, with a 14 T PPMS system. The home-made setup is tailored for dielectric constant measurements in a capacitor-like arrangement. It gives the possibility to select between an AH 2700A Ultra-precision Capacitance Bridge, for relatively low-frequencies (50 Hz - 20 kHz) or a Solatron 1260 Impedance/Gain Phase Analyser, for the high-frequency region up to 32 MHz; the latter is being used together with a 1296A Dielectric Interface System in order to cope with ultra-low capacitance levels. A Lakeshore 370 temperature controller was utilized to cover a broad temperature range (5 ≤T≤ 180 K).

## III. RESULTS AND DISCUSSION

### A. Macroscopic Properties.

The temperature dependent magnetic susceptibility, $\chi(T)$, of the different NaMnO$_2$ polymorphs, qualitatively appears similar, with exception of the presence of a broad maximum (~200 K for α–polytype),[36] which apparently shifts to higher temperature in the β–phase (Figure 1a). Such a broad feature is a general characteristic of low-dimensional antiferromagnetic systems. However, from $\chi(T)$ data alone no evidence for a transition to a long-range ordered state is observed.

On the other hand, the heat capacity, $C(T)$, measured in zero magnetic field displays several very weak anomalies (Fig. 1), possibly of magnetic origin. In order to highlight these features, we first estimated the phonon contribution to the specific heat, $C_{ph}(T)$, and subtracted it from the experimentally measured heat capacity. Here, $C_{ph}(T)$ assumes a sum of Debye contributions ($2 < T < 280$ K), following the procedure used before for other low-dimensional spin systems: [46,47]

$$C_{ph}(T) = 9R \sum_{i=1}^{2} C_i \left(\frac{T}{\theta_D^{(i)}}\right)^3 \int_0^{x_D^{(i)}} \frac{x^4 e^x}{(e^x-1)^2} dx \qquad (1)$$

with $R$ (8.314 J mol$^{-1}$ K$^{-1}$) the gas constant, $\theta_D^{(i)}$ the Debye temperature and $x_D^{(i)} = \theta_D^{(i)}/T$, while fitting was based on an optimization approach using the minimum number of free parameters. In our case, the $C_{ph}(T)$ was approximated by two Debye functions, addressing the relatively different atomic masses of the constituent element-coupled vibrations (cf. Na-O and Mn-O) in the β-NaMnO$_2$. This yielded the fitting parameters, $C_1= 0.55(2)$, $C_2= 2.0(2)$ and $\theta_D^{(1)} = 287(22)$ K, $\theta_D^{(2)} = 510(15)$ K (Figure 1a). The vanishingly small magnitude of $C(T)$ at very low temperatures, in accord with the β-phase insulating nature, agrees well with the $\propto T^3$ term that corresponds to phonons (eq. (1)).

The outcome of the subtraction of $C_{ph}(T)$ from the total heat capacity is shown in Figure 1b. As the corresponding anomalies in the differential $C(T)$ are very small, pointing to some sensitivity to the defects in the lattice structure (*vide-infra*), and the estimated phonon part uncertainties are high, they render further analysis to assess the differential $C(T)$ as a likely magnetic contribution, $\Delta S_{mag} = \int \frac{C_{mag}(T)}{T} dT$, unfavorable at this stage. The identification, though, of the two fairly broad humps centered at ~95 K ($T_{N2}$) and ~200 K ($T_{N1}$), would suggest that β-NaMnO$_2$ undergoes two transitions. These qualitative $C(T)$ characteristics therefore require further study to inquire the role of magnetic interactions in such phase changes.

### B. $^{23}$Na NMR Dynamics near the Transitions.

A critical aspect of many macroscopic thermodynamic properties is the role of the material's microscopic dynamical response. Techniques capable of detecting spin dynamics on a local scale, such as solid-state $^{23}$Na NMR, can therefore be helpful to understand the complex behavior of β-NaMnO$_2$. The $^{23}$Na NMR powder spectra of β-NaMnO$_2$ were measured between room temperature and 50 K, where they become very broad and, consequently, the signal becomes very weak and difficult to measure (Figure 2a). At 300 K, the spectrum has a characteristic powder line shape for a quadrupole $I= 3/2$ nuclei with the quadrupole asymmetry parameter of $\eta \approx 0$. A closer inspection of the satellite ($\pm 3/2 \leftrightarrow \pm 1/2$) transitions of the 300 K spectrum terminated around $\pm v_Q = \pm 1.28$ MHz from the narrow central transition ($1/2 \leftrightarrow -1/2$) line (upper inset to Figure 2a) shows that the expected singularity is rounded, which is consistent with a high-degree of Na local site disorder. Here $v_Q$ is the $^{23}$Na quadrupole frequency. On cooling below $T_{N1}$, there is almost no change of the central transition line. However, a close inspection of the $^{23}$Na NMR satellite line reveals that a shoulder starts to gradually broaden well beyond $\pm\nu_Q$. This is clearly seen as a growth of the NMR signal intensity on both sides of the satellite shoulder (lower inset of Figure 2a). As the positions of the satellite shoulder remain nearly at the same frequency, the quadrupole frequency must also remain the same through the transition at $T_{N1}$. This suggests that no structural deformation takes place in the vicinity of the Na site, corroborating that the high temperature transition ($T_{N1}$) is magnetic in origin. Moreover, below $T_{N1}$ the intensity of the sharp central peak multiplied by temperature (to counterbalance the changing Boltzmann population), starts to progressively decrease with decreasing temperature below $T_{N1}$ (Figure 2b). The broadening of the NMR line beyond the satellites can thus be attributed to growing internal magnetic fields at certain Mn ion sites, while the gradual wipeout of the central line below $T_{N1}$ (Figure 2b) reveals that the high-temperature paramagnetic-like signal vanishes only gradually, as it remains present at all temperatures below $T_{N1}$. This leads us to the important conclusion that the magnetic state below $T_{N1}$ is inhomogeneous. On further cooling below $T_{N2}$, the $^{23}$Na NMR line shape broadening becomes really pronounced as the spectrum becomes completely dominated by the broad distribution of internal (hyperfine) magnetic fields and the sharp central peak almost disappears. These line shape changes verify that β-NaMnO$_2$ indeed undergoes two successive transitions to magnetically ordered states, at ~200 K and ~95 K, in agreement with the assignment of subtle peaks in the differential $C(T)$ as magnetic transitions (Fig. 1b).

Additional information about the two magnetic transitions is deduced from the $^{23}$Na spin-lattice relaxation rate $1/T_1$, which was determined from fitting of $^{23}$Na magnetization recovery curves (Figure 3a) to the magnetic-relaxation model for $I=3/2$,[48]

$$M(t) = M_0 \left[1 - s\left(1/10 e^{-\left(\frac{t}{T_1}\right)^a} + 9/10 e^{-\left(\frac{6t}{T_1}\right)^a}\right)\right]. \qquad (2)$$

Here, $s < 1$ accounts for imperfect inversion of $^{23}$Na nuclear magnetization after the initial π pulse, while $\alpha$ stands for a stretching exponent. In the high-temperature paramagnetic (PM) regime, $1/T_1$ is nearly temperature independent, $1/T_1 = 35(1)$ s$^{-1}$ (Figure 3b). Such temperature independence is in fact anticipated for an exchange-coupled antiferromagnetic (AFM) insulator in the paramagnetic phase. The stretching exponent is $\alpha = 0.88$ (Figure 3a); a value slightly below 1 implying a small distribution of relaxation rates expected in experiments on powder samples. The transition to the magnetic state at $T_{N1}$ is accompanied by a sizeable step-like increase in the $1/T_1$ value to $1/T_1 = 66(5)$ s$^{-1}$ and a gradual reduction of the stretching exponent (Figure 3a). The latter indicates that the distribution of the spin-lattice relaxation times suddenly starts increasing below $T_{N1}$ thus indicating growing magnetic inhomogeneity between $T_{N1}$ and $T_{N2}$ which is in accord with the line shape changes (Figure 2a). In fact, as two-step magnetization-recovery curves are clearly observed below $T_{N1}$ (e.g., measurement taken at 100 K shown in Figure 3a), the fit of the magnetization recovery curves in the $T_1$ experiment is significantly improved if two relaxation components are included. Here, the relative intensity of one of the components (AFM1) increases at the expense of the second PM component, the latter in close analogy to the wipeout effect of the narrow central line (Figure 2b).

We stress that no obvious critical fluctuations leading to diverging $1/T_1$ could be detected at $T_{N1}$. The likely reason is the nature of magnetic fluctuations, which according to the expression, $\frac{1}{T_1} = \frac{2\gamma_n^2 k_B T}{(\gamma_e \hbar)^2} \sum_{q^\rho} A_q^\rho A_{-q}^\rho \frac{\chi_\perp''(q^\rho,\omega)}{\omega}$ (where $A_q$ denotes the hyperfine coupling of the $^{23}$Na nuclei with the electronic magnetic moments, $\chi''$ is the imaginary part of the dynamical susceptibility and ω is the Larmor frequency), could be filtered out in the $1/T_1$ measurements for highly symmetric Na (octahedral) sites. On the other hand, on approaching the lower transition temperature at $T_{N2}$, the $1/T_1$ of the paramagnetic PM component is rapidly enhanced, suggesting the onset of critical fluctuations. A phenomenological fit of the critical model $1/T_1 = A + B(T-T_{N2})^{-p}$ to the PM data in the temperature range between $T_{N2}$ and 110 K, yields the critical exponent $p = 0.45(10)$ for $A = 66(5)$ s$^{-1}$ and $T_{N2} = 95.0(5)$ K (Figure 3b). Such critical enhancement demonstrates that the magnetic fluctuations that govern the transition at $T_{N2}$ cannot be filtered out anymore at the Na site. This is also consistent with the observed dramatic $^{23}$Na NMR line shape changes (Figure 2a). The temperature dependence of the other component (AFM1), which we attribute to the already magnetically ordered regions in the sample, is much more subtle (Figure 3b). Finally, at $T < T_{N2}$, the two components in the magnetization recovery curves are not obvious any more (Figure 3a), so we resort back to a single-exponential fit (eq. 2). However, a very low stretching exponent of 0.34 has to be employed. Such a strikingly low value of $\alpha$ indicates an extremely broad distribution of relaxation times, hence a broad distribution of local magnetic environments below $T_{N2}$. At the same time $1/T_1$ is strongly suppressed below $T_{N2}$ and exhibits an activated type of dependence ($1/T_1 \propto T^2 \exp(-\Delta/T)$, Figure 3b), indicating the opening of an excitation gap, $\Delta$, in the low-temperature phase.

## C. Crystallographic structure

Critical to understanding such transformations is the way magnetic ions are arranged in the underlying lattice structure that establishes nearest-neighbor exchange terms and stabilizes non-degenerate ground states. High quality data collected on the WISH diffractometer enables the analysis of the crystallographic structure of β-NaMnO$_2$. The main reflections of the neutron powder diffraction (NPD) pattern are consistent with the *Pmmn* space group, with cell parameters $a_o = 4.7851(2)$ Å, $b_o = 2.8570(8)$ Å, $c_o = 6.3287(4)$ Å, at 300 K. The Rietveld refinement of the main nuclear reflections (300 K), with the *Pmmn* model [29] (Mn1 in 2b position z= 0.617(5), Na1 in 2b z= 0.125(4), O1 in 2a z= 0.365(6) and O2 in 2a z= 0.872(6)), suggests a significant degree of "anti-site" defects between the Mn and Na sites that leads to an average occupation of ~80:20 (see Figure S2, Supplemental Material) [28]. Moreover, the refinement points to an unexpectedly large-value for the oxygen thermal parameter ($U_{iso} \sim 0.038(2)$ Å$^2$). The use of anisotropic temperature factors in the refinement results in a clear elongation of the thermal ellipsoids along the c-direction (see Figure S2) [27] indicating strong positional disorder. Following this suggestion we spilt the two oxygen positions along the c-axis and the refinement converged to a splitting of ~0.5 Å and ~70:30 occupancy of the resultant sites, with normal isotropic thermal parameters ($U_{iso} \sim 0.015(2)$ Å$^2$). It is worth stressing that the split and especially the occupancy of O1 and O2 resemble the anti-site occupancy of the Mn and Na atoms; in particular, as shown in Figure S2 this distortion is needed to satisfy the coordination requirements of the Na and Mn cations.

A crucial feature of the 300 K NPD pattern, in association with the above analysis, is the presence of additional reflections that could be ascribed to a nuclear modulation (Figure 4). In support to this comes earlier transmission electron microscopy (TEM) work on β-NaMnO$_2$ [33], where it was pointed out that formation of planar defects establishes short-ranged ordered regions that locally (i.e. on the length scale of a few unit cells) follow the stacking sequence of NaMnO$_2$ layers characteristic of either the α- or the β- phases. Importantly, long-period stacking sequences, with a modulation vector $q=(\alpha 00)$ $\alpha \approx 0.1$ (consistent with the cell choice reported in the present work), were also required for indexing the additional satellite peaks observed in both electron and synchrotron X-ray diffraction data. From a LeBail fit of the WISH data we obtained an optimal modulation vector $q = (0.077(1),0,0)$, accounting for satellites up to the second order in the NPD pattern. Some small satellite reflections, however, are sliding off the calculated position (Fig. 4), suggesting that the other two components of the modulation vector may be slightly different from zero. Refinements where the other two components of $q$ were allowed to vary proved unstable and did not lead to reasonable results. The obtained value of $q$ is near the commensurate 1/13 position, which explains why the 1/6 value used before in the synchrotron X-ray diffraction patterns indexed well a large number of satellite peaks.

The observation of the satellite reflections in both NPD and TEM measurements and the refinement of the average nuclear structure, indicate the possibility of a compositional modulation in the structure that can be modelled through the superspace formalism. [49, 50] The theory of (3 + D) superspace groups, introduced by de Wolff (1974, 1977), [51, 52] is widely used to describe the symmetry of commensurate and incommensurate modulated structures. In order to understand the NPD pattern of WISH we therefore used a (3+1)-dimensional superspace approach considering an occupational modulation for all the sites in the average nuclear structure. In order to derive the possible superspace groups we performed group symmetry analysis with the help of the ISODISTORT Software Suite [44] starting from the refined average structure and the propagation vector $q=(\alpha 00)$. Having taken into account the observed reflection conditions and the symmetry properties of the modulation vector, the symmetry analysis led to the $Pmmn(\alpha 00)000$ superspace group as the best solution, corresponding to the $\Sigma_1$ irreducible representations (IRs), with order parameter direction (OPD) $P(\sigma,0)$. [53]

To account for the compositional modulation a step-like (Crenel) function is introduced for every site in the structure. The Crenel function is defined as [54]

$$p(x_4) = 1 \in \langle x_4^0 - \Delta/2, x_4^0 + \Delta/2 \rangle \\ p(x_4) = 0 \notin \langle x_4^0 - \Delta/2, x_4^0 + \Delta/2 \rangle \quad (3)$$

where $x_4$ is the internal (fourth) coordinate in the (3+1)D approach and $\Delta$ is the width of the occupational domain centered at $x^0_4$ ($\Delta$ corresponds also to the average fractional occupancy of the site). The modulation functions on the same cation site are constrained to be complementary, meaning that in every point of the crystal the site is occupied (this results in the equations $\Delta[Mn_i]+\Delta[Na_i]=1$ and $x_4[Mn_i]=1-x_4[Na_i]$ for each cation site). For the split oxygen positions we introduce a similar constraint, imposing that in any position in the crystal we have the superposition of the two split sites. Regarding the origin along the fourth axis, the superspace group constrains this value to two equivalent values: 0 and 0.5, thus making the choice trivial. Moreover, an additional constraint is introduced regarding the two Mn/Na sites. The electron diffraction measurements, reported by Abakumov et al. [33] suggest that the quasi-periodic stacking sequences of the $NaMnO_2$ layers entail coherent stacking faults, a feature which points that their modeling can be reduced to the alternation sequence of the Na and Mn cations. We followed a similar approach for the modeling of the NPD pattern assuming that the step-like functions were constrained to have in every $NaMnO_2$ plane the right Mn/Na ordering, that is to say, when one site switches from Mn to Na the other changes from Na to Mn. The crystallographic model built in this way was employed for qualitative Rietveld refinements. Broad, asymmetric reflections throughout the NPD pattern, mainly due to defects (e.g. stacking faults) and strain make such analysis hard to optimize, raising the agreement factors and making a quantitative refinement difficult. The Rietveld plot, over a wide d-spacing range, is shown in Figure 4 and the associated reliability factors are, $R_p$= 8.81%, $R_{wp}$= 12.73%, $R_{main}$= 9.96%, $R_{sat\pm 1}$= 15.41%, $R_{sat\pm 2}$= 14.79%. Despite the apparent reflection broadening, our model shows good agreement for the modulated parts of the profile, especially obvious in the relatively short *d*-spacing region of the pattern (inset in Figure 4). The crystallographic parameters of the compositionally modulated β-$NaMnO_2$ at 300 K, on the basis of a (3+1)D Rietveld analysis with the $Pmmn(\alpha 00)000$ superspace group (a= 4.7852(4) Å, b= 2.85701(8) Å, c= 6.3288(4) Å, *α=0.077 (1)*) are compiled in Table S1 [28].

This single-phase structural model, despite the presence of low intensity reflections ascribable to a small amount of the α-phase and MnO (Fig. 4), takes into account almost all the satellites present in the NPD pattern of the β-phase, as compared to the two-phase description on the basis of the $B2/m(\alpha\beta 0)00$ superspace group derived before from the analysis of the synchrotron X-ray powder diffraction data [33]. The nuclear structure model obtained here is shown in Figure 5. This is consistent with the one proposed by Abakumov et al.,[33] entailing coherent intergrowth of stacking sequences of $NaMnO_2$ layers along the $a_0$-axis, characteristic of the α- and β- polytypes. It may be considered as good approximation to the real chemical phase, as planar defects, seen by electron microscopy, could violate the idealized

Crenel-type function used in the present analysis of the NPD data. In this model, the MnO$_6$ octahedra throughout the structure display strong Jahn-Teller distortion (see Figure S3 [28], for oxygen-cation distances in the (3+1)D approach), with four short bonds below 2 Å and two long ones around 2.4 Å, in a fashion analogous to the α-NaMnO$_2$ [36]. On the other hand, while Na is also octahedrally coordinated to oxygen, the distances involved are longer due to its larger ionic radius. Moreover, in an effort to visualize the degree of compositional modulation in the β-NaMnO$_2$ structure, Fourier maps of the observed structure factor (Figure 6) involving the atomic sites in the $zx_4$-plane were computed on the basis of the observed NPD intensities and the calculated phases. Figure 6a shows the complementary occupation of the cation sites without any particular modulation of the z coordinate. On the contrary, from the Fourier maps centered at the oxygen positions (Figure 6b) it is inferred that the site-splitting observed in the average structure is needed in order to satisfy the coordination requirement of the Mn$^{3+}$ Jahn-Teller active cation. In fact, it is noted that when the Na and Mn swap sites (cf. compositional modulation), the same happens in the oxygen split positions so that the bonding requirements are restored as depicted in Figure S3 [28]. Our approach demonstrates that having taken advantage of the superspace formalism to describe the compositional modulation of the Mn and Na sites in a single-phase atomic configuration, the incommensurate β-NaMnO$_2$ structure can be depicted as a coherent intergrowth of two types of NaMnO$_2$ layers, reflecting the α- and β- polytype oxygen coordinations (Fig. 5).

**D. Magnetic Structure Evolution.**

In view of the complex nuclear modulated structure observed in the NPD profiles of β-NaMnO$_2$ it is challenging to evaluate the correlation between the crystal and magnetic structures as the sample temperature is lowered. The temperature evolution of the diffraction pattern demonstrates the presence of two magnetic transitions (Figure 7).

First, below $T_{N1}$~ 200 K there is an intensity increase at magnetic Bragg peak positions corresponding to a propagation vector **k**=(½ ½ ½) with respect to the *Pmmn* orthorhombic average structure. These reflections grow quickly below the magnetic transition temperature and their broad Lorentzian-like profile is an indication that the magnetic domain is sensitive to the strain and defects present in the nuclear structure (refer to Figure 6), complying with the broadening of $^{23}$Na NMR spectra (inset, Figure 2a). Moreover below about 100 K the diffraction patterns show the development of additional reflections (Figure 7a). This new set of peaks can be indexed assuming the combination of the magnetic propagation vector **k** and the nuclear one **q**, giving magnetic intensity at the positions hkl±**[k±q]**. It is worth noting that the temperature dependence of the integrated intensity (Figure 7b) of these two sets of reflections possesses different critical behavior, thus suggesting that the two magnetic orders likely fall into different universality classes. In particular the fit of the ½½½ reflection with power law $I = I_0 [1-(T/T_N)]^{2\beta}$ gives a critical exponent of β= 0.33(4), indicating interactions of a 3D nature, instead, the **k±q** satellites possess an exponent of β= 0.15(8), which is more consistent with 2D interactions (Figure 7b). Careful analysis of the diffraction pattern reveals the presence of some additional low intensity reflections that are not indexed with the previous propagation vectors. These extra reflections are ascribed to a small-content of MnO impurity and the α-polymorph.

Let us first discuss the important changes in the NPD pattern that were observed below 200 K. In order to establish the possible magnetic space group we performed magnetic symmetry analysis with the help of the ISODISTORT software. [44] The NPD patterns show that no clear magnetic intensity is observed on the nuclear satellite reflections, therefore pointing that the magnetic structure is not strongly related to the nuclear modulation at least in the 100 < *T*< 200 K temperature range. For this reason, magnetic symmetry analysis was initiated on the basis of parent average *Pmmn* nuclear structure (Figure S1, Table S1) [28] and the propagation vector **k**= (½ ½ ½). The results of the symmetry analysis are reported in Table S2 [28]. The best agreement between observed and calculated patterns was obtained for the mR1 representation, with order parameter direction (OPD) *P1(a,0)*, corresponding to the magnetic space group *$C_a2/c$,* with a change in the unit cell with respect to the parent structure described by the transformation matrix {(0,-2,0),(0,0,2),(-1,1,0)}. It is worth underlining that the space group *$C_a2$* also gives a reasonably good result (Table S2), but with an increased number of refinable variables, thus suggesting the higher symmetry option *$C_a2/c$* as the best solution. Combining the mR1 *P1(a,0)* IRs with the compositional modulated structure, the *$C_a2'/c'(a0\gamma)00$* magnetic superspace group is obtained. With the latter we then carried out Rietveld refinements, with the representative 100 K profile. The Rietveld plot is shown in Figure 8, and the refined parameters are compiled in Table S3 [28]. The associated reliability factors are, $R_{Fobs}$= 8.46% for the nuclear reflections and $R_{Fmag}$= 12.50% for the magnetic ones, while the $R_P$= 13.88%. Their values are rather on the high side, due to pronounced *hkl*-dependent broadening, likely arising from the presence of planar defects. The magnetic structure is drawn in Figure 9, projected in the same plane as the nuclear one (Figure 5, top panel). It entails antiferromagnetically coupled Mn-chains running down the **$b_o$**-axis (**$a_o$**, **$b_o$** and **$c_o$** setting is with respect to the orthorhombic *Pmmn* unit cell), stacked in a zig-zag fashion when viewed in an $a_0c_0$-plane projection (Figure 9a) that gives rise to antiferromagnetically coupled, corrugated $MnO_2$ layers (Figure 9b). A similar collinear spin-model has been utilized before for the description of the magnetic state in the isomorphous β-$LiMnO_2$, where three-dimensional long-range order is established at $T_N$~ 260 K.[55]

The derived spin-configuration for β-$NaMnO_2$, though, indicates a commensurate ordering only for the Mn2-site, as a similar ordering on the Mn1 site would have generated strong magnetic intensity at the nuclear satellite reflections, a case that is not supported by the NPD data. In this compositionally modulated nuclear structure, between 100 < *T*< 200 K only the $NaMnO_2$ layer stacking sequences characteristic of the β- polytype carry a net magnetic moment. Such a magnetically inhomogeneous state is consistent with the wipeout of the central $^{23}$Na-NMR line (Fig. 2b) and the two-component nuclear spin-lattice relaxation in the same temperature range. The magnetic moment of Mn2 sites has been computed as µ≅ 2.38(10) µ$_B$ at 100 K, but as the observed NPD profile shows fairly broad magnetic peaks, the attained staggered moment may be an underestimate (cf. the full moment for spin-2 $Mn^{3+}$ is expected to be 4 µ$_B$).

When temperature is lowered below $T_{N2}$~ 100 K, the incommensurate-like magnetic ordering appears to be described with a combination of the magnetic, **k**, and nuclear, **q**, propagation vectors suggesting that the second transition takes place because longer-range magnetic correlations are established in the alpha-like stacking sequence(s). Assuming that the same superspace group defines also the magnetic order at *T*< 100 K and taking into account a Mn1-site spin-configuration similar to that of the Mn2-site, magnetic scattering is calculated only for the **k+q** satellite positions. However, its relative intensity does not match the experimentally observed one, pointing out that additional spin modulation of the existing structure is required in order to adequately reproduce the observed magnetic NPD pattern. Rietveld refinements of the magnetic structure confirmed that the magnetic phase below $T_{N2}$ can be described by a proper screw component, with propagation vector **k+q** for both Mn1 and Mn2 sites, while refinements assuming a spin-density wave type of structure produced worse agreement factors and unphysical moment size for the Mn1 site. The corresponding Rietveld refined 5 K NPD profile is shown in Figure 10, with the refined magnetic parameters compiled in Table S4 [27]. The associated reliability parameters are, $R_{Fobs}$= 8.41% for the nuclear reflections and $R_{Fmag}$= 9.4% for the magnetic ones, while the $R_P$= 16.6% is relatively poor again due to the extreme peak broadening. The magnetic structure below $T_{N2}$ is depicted in Figure 11a-b.

To a first approximation the spin configuration is similar to the commensurate one that develops below $T_{N1}$, but at the "boundary" of the α- and β- like stacking sequences (Figure 5), as the ordering at the Mn1-site (α-$NaMnO_2$ layer stacking sequence) acts as a perturbation to the Mn2-site, the Mn-spins start to rotate away from the commensurate structure type (Figure 9a). Within this modulated behavior, the $Mn^{3+}$ magnetic moment takes the lowest values within the $NaMnO_2$ layers characteristic of the α-polytype (likely due to their higher degree of spin-frustration), while it grows in magnitude as we move within the β-like stacking sequences, reaching a maximum, $\mu \cong 3.5(10)$ $\mu_B$, at their mid-point (see Figure S4) [28]. Such a non-trivial magnetic order is in line with very broad distribution of spin-lattice relaxation times found by NMR below $T_{N2}$ (cf. low-value of the stretching exponent), implying a broad distribution of local environments. This complexity might be an outcome of the system's effort to relieve competing interactions amongst neighboring spins in the β-$NaMnO_2$ modulated nuclear structure, therefore requiring further insights on the role of geometric frustration.

**E. Parameterization of Magnetic Excitations.**

Since the NPD and the NMR resolved two magnetic regimes, the magnetic fluctuations of β-$NaMnO_2$ were studied by inelastic neutron scattering (INS). An overview of the measured INS response, well within the magnetically ordered state (5 K), is shown in Figure 12a for experiments on the MARI spectrometer. A complementary insight on the low energy magnetic dynamics was offered with higher resolution through the DCS spectrometer (Figure 13). At low temperatures (1.5 K and 75 K) the DCS spectra show clearly the presence of a spin-gap in the excitation spectrum, with little change in the gap energy, $\Delta \sim 5$ meV. A pronounced change is observed at 100 K with a filling of the gap, yet with the presence of significant magnetic scattering even at $T > T_{N1}$ (see Figure S5) [28].

As the measured neutron scattering cross section is proportional to the structure factor $S(\vec{Q}, \hbar\omega)$, for a powder material, the measured, momentum integrated neutron intensity is proportional to the following average at a fixed $|\vec{Q}|$, $\bar{I}(\vec{Q}, \hbar\omega) = \int \frac{d\Omega \, S(\vec{Q}, \hbar\omega)}{d\Omega \, Q^2}$. Obtaining microscopic exchange interactions that form the basis of the magnetic Hamiltonian from powder neutron data is rather difficult owing to the averaging over all reciprocal space directions, $|\vec{Q}|$. However, applying sum rules allows information to be obtained about the interactions and correlations in a general way which is independent from the microscopic Hamiltonian. We outline this method in the following.

In the absence of a full theory for the magnetic exchange interactions in β-$NaMnO_2$, and lack of single crystal data, we have parameterized the dispersion $E(\vec{Q})$ with a phenomenological expression which satisfies the periodicity of the lattice and hence Bloch's theorem. One possible form of the dispersion, consistent with lattice periodicity can be written as a Fourier series $E^2(\vec{Q}) = \Delta + \sum_d B_d \sin^2(\vec{Q} \cdot \vec{d})$, where $\vec{d}$ is a bond vector connecting nearest-neighbor (NN) spins and $B_d$ are coefficients in this series expansion, and Δ is the magnitude of the spin-gap. Because the magnetic excitations appear relatively sharp in energy (Figures 12, 13), we could utilize the single mode approximation (SMA) which states that the structure factor, which is proportional to the neutron cross section, is dominated by a single resonant mode.

The problem of deriving a parameterization of the neutron cross section, $S^{\alpha\alpha}(\vec{Q}, \hbar\omega) = S(\vec{Q}) \, \delta(\hbar\omega - E(\vec{Q}))$ (delta function being numerically approximated by a Lorentzian with the energy resolution width), is reduced to finding an expression for $S(\vec{Q})$. To do this, we apply the Hohenberg-Brinkmann first moment sum rule, [56] which applies to the case of isotropic exchange and is closely related to the ground state magnetic energy. Effectively the first moment sum relates $S(\vec{Q})$ to the dispersion $E(\vec{Q})$ through the following expression:

$$S(\vec{Q}) = \frac{\hbar \langle \omega \rangle}{E(\vec{Q})} = -\frac{1}{3} \frac{1}{E(\vec{Q})} \sum_{\vec{d}} J_d \langle \vec{S}_0 \cdot \vec{S}_d \rangle \left(1 - \cos(\vec{Q} \cdot \vec{d})\right) \qquad (4)$$

In view of this, the single-mode approximation and parameterization of the dispersion, $E(\vec{Q})$, allows us to characterize which correlations are important and also determine the dimensionality of the excitations. In particular, the energy gap in a powder averaged constant-$Q$ scan is sensitive to the dimensionality of the interactions. This fact was previously used to show that α-NaMnO$_2$ is dominated by one-dimensional magnetic correlations.[37]

Comparison of the powder averaged spectra for β-NaMnO$_2$ against its closely related α-NaMnO$_2$ system (see Figure S6)[28] points to several key differences. First, the spectral weight in α-NaMnO$_2$ is concentrated at low energies near the energy gap edge, while it is much more evenly distributed in energy in the case of the β-NaMnO$_2$ variant. The scattering is also much more strongly peaked[56] in momentum for β-NaMnO$_2$, which is indicative of the higher (cf. than the quasi-1D of the α-phase) dimensionality of the associated spin correlations. In addition, considerable spectral weight is located at the top of the excitation band and the scattering is much more well-defined in momentum than in the α-polytype. Such qualitative observations, suggest that β-NaMnO$_2$ may be more two-dimensional than the α-phase. We have therefore simulated the powder averaged spectra by considering the case of the two-dimensional spin-exchange, with dominant correlations along the **b$_0$**-crystal axis. We have taken the dispersion relation to have the following phenomenological expression:

$$E^2(\vec{Q}) = B_0 + B_1 sin^2(\pi K) + B_2 sin^2(\pi H) + \cdots + B_3[sin^2(\pi(K+H)) + sin^2(\pi(K-H))] \quad (5)$$

which is consistent with the periodicity of the lattice (*Pmmn* symmetry) and gives a minimum at half integer positions, relating the observed magnetic Bragg peaks. We have chosen $B_0 = 25$ meV$^2$ to account for the spin-gap (Δ), $B_1 = B_2 = 625$ meV$^2$ and $B_3 = 400$ meV$^2$.

To extract an estimate for the exchange constants, we have put the inelastic magnetic response on an absolute scale using the internal incoherent elastic line as a reference. The absolute calibration combined with the first moment sum rule afforded an estimate of $J_d \langle S_0 \cdot S_d \rangle$. Combined with the collinear magnetic structure, we have estimated a strong exchange along the **b$_0$**, $J_1 = 5.0 \pm 1.0$ meV and a weaker one along **a$_0$**, $J_3 = 1.5 \pm 1.0$ meV (Figure S1)[28].

The total integrated spectral weight (elastic and inelastic) is constrained by the zero$^{th}$ moment sum rule which can be summarized as follows:

$$\frac{\int d^3q S(\vec{Q},\hbar\omega)}{\int d^3q} = S(S+1). \quad (6)$$

Integrating the INS data by using the elastic incoherent scattering of the vanadium as an internal standard gives the inelastic contribution to the above integral being 1.8(3). Including the ordered moment in the elastic channel and noting that there are two Mn$^{3+}$ ions per unit cell gives a total integral of 4.7(4) for this sum. Given the expected value for S= 2 is 12, this indicates that more than half of total moment resides elsewhere in momentum and energy. One possibility is for a large fraction residing in diffuse scattering, which maybe resulting in a low-energy contribution that is beyond the resolution of the spectrometer, while it is in agreement with the broad shape of the magnetic reflections in the diffraction data and with the high density of structural defects present in the material.

**F. Incommensurate Structure and Frustration.**

We have seen that the magnetic long-range order of β-NaMnO$_2$ is strongly correlated with its structural complexity, which is established through the relief of frustration. Importantly, competing interactions between spins and their complex magnetic orders are known to motivate spectacular cross-coupling effects that lead to improper ferroelectricity in frustrated magnets.[57] Establishing cross-control of the magnetic and ferroelectric polarisations challenges scientific endeavours as striking new multi-ferroic device concepts may be realized.[58] A key question then is whether the compositionally modulated nu-clear structure and magnetic order in β-NaMnO$_2$ may also stimulate competing degrees of freedom that can become cross-correlated through the symmetries[59] of the associated magnetic and nuclear orderings. Preliminary evidence for such a type of behavior in β-NaMnO$_2$ was first reported by Bakaimi et al. who demonstrated that the temperature-dependent dielectric permittivity, ε′(T), displays two small anomalies, near the $T_{N1}$ and $T_{N2}$ transitions discussed here.[60] Since the explanation of possible magnetoelectric coupling needs the understanding of the crystal and magnetic symmetries, these early findings remained unexplored. Now that these structures are known, through the current work, it is worth revisiting the coupling of the afore-mentioned properties.

Let us now glance through the dielectric response of β-NaMnO$_2$ and compare it to that of α-NaMnO$_2$. Bearing in mind that the magnitude of the dielectric permittivity anomalies in β-NaMnO$_2$ becomes larger with the application of an intense electric field,[60] here instead we utilized a progressively stronger external magnetic field, hoping for enhanced changes in the ε′(T). Our dielectric permittivity experiments, however, identified only small anomalies in ε′(T, H) curves that coincide with the onset of antiferromagnetic orders taking place in the bulk α- ($T_N$= 45 K) and β- ($T_{N2}$= 95 K) phases. In β-NaMnO$_2$, no other low-temperature ε′(T,H) signature is observed that could indicate contributions from α- and β-type structural domains, as local probes have resolved before.[41] Moreover, the magnetoelectric coupling must be weak in both NaMnO$_2$ materials, as very little changes are brought about despite the strength of the externally applied magnetic field (Fig. 14). Having taken into account the symmetry-imposed constraints for the free-energy [61] in the α- and β- magnetic phases, it is conferred that the spatial inversion symmetry is not violated, excluding the possibility of improper ferroelectricity in the magnetically ordered states (see Section S7, Supplemental Material). In this respect, it is postulated that the observed small anomalies in the dielectric constant are likely related to the non-linear, higher order terms (e.g. biquadratic term ~$E^2H^2$) that are operative in chemically diverse systems, ranging from planar magnets[62, 63] and three-dimensional magnetoelectric perovskites (AMnO$_3$, A= Y, Bi)[64, 65] to quantum para-electrics (EuMeO$_3$)[66, 67].

## IV. SUMMARY & CONCLUSIONS.

The present work entails a thorough study of the crystallographic and dynamical properties of the β-NaMnO$_2$. The proposed single-phase nuclear structure model, takes advantage of the superspace formalism to describe the incommensurate compositional modulation (propagation vector, **q**= (0.077(1), 0, 0)) of the Mn and Na sites that can be depicted as an intergrowth α- and β- like oxygen coordinations. This peculiar topology strongly influences the physical and chemical properties of the material and underlines the role of the nearly degenerate in energy α and β layer stacking sequences. The remarkable flexibility of β-NaMnO$_2$ to adapt its lattice topology is likely at the basis of the particular high charge capacity of the system as a Na-ion cathode material,[35] but also may corroborate to the stability of the various non-stoichiometric phases[41] accessible through its electrochemical Na-intercalation/removal.[68]

Moreover, the magnetic structure of β-NaMnO$_2$ was solved on the basis of time-of-flight neutron powder diffraction data and found to be strongly mediated by the material's inherent lattice topology. First, below $T_{N1}$ (200 K), a collinear commensurate antiferromagnetic state, involving only the β-like stacking sequences, develops with a propagation vector **k**= (½ ½ ½). Then, a second magnetic transition is observed at $T_{N2}$ (95 K), marked by new satellite reflections ascribed to the interaction of **k** with the compositional modulation vector **q**. The new magnetic ordering is due to the relief of the magnetic frustration in the α-like sheets that in turn influences the ordering in the β-like stacking sequences, and instigates a cooperative proper-screw magnetic state. Here, the lattice topology of the Jahn-Teller active Mn$^{3+}$ cation drives the original 3D spin correlations (T< $T_{N1}$) to become 2D in character. Inelastic neu-tron scattering and $^{23}$Na NMR provide evidence that a spin-gap ($\Delta$= 5 meV) opens in the excitation spectra, in line with the 2D nature of the magnetic interactions at $T< T_{N2}$.

Overall, structure and dynamics point that the incommensurate β-NaMnO$_2$ structure can relay a magnetocapacitance effect in the low-temperature magnetic state. Such a structural complexity, inquires whether controlled engineering of coherent defects may impart the material with novel technological capabilities. In view of this, it is worth considering that in the compositionally modulated β-NaMnO$_2$, domain-wall (DW)-like phenomena [69] associated with the abundance of the α- and β- interfaces (Figures 6 and 12), rather than extended domains themselves, may be the active element in promoting some degree of topologically correlated (related to DW), cooperative magnetic and electric dipole arrangements. The way electronic structure changes at such interfacial regions could be relevant in order to manipulate the magnetoelectric response [70] even in this class of non-perovskite compounds and warrants further exploration.

## ACKNOWLEDGMENTS

We thank the Science and Technology Council (STFC) for the provision of neutron beam time at ISIS Facility. Access to DCS was provided by the Center for High Resolution Neutron Scattering, a partnership between the National Institute of Standards and Technology and the National Science Foundation under Agreement No. DMR-1508249. This work was partly funded by the Carnegie Trust for the Universities of Scotland, the Royal Society, and the EPSRC. Partial funding was also secured through the framework of the Heracleitus II project (Grant No.349309.WP1.56) co-financed by the Ministry of Education and Religious Affairs, Greece and the European Social Fund, European Union (Operational Program 'Education and Lifelong Learning' of the National Strategic Reference Framework, NSRF, 2007-2013).

## AUTHOR INFORMATION

**Corresponding Author**

\* e-mail: lappas@iesl.forth.gr

**Present Address**

† Ioanna Bakaimi, Department of Chemistry, University of Southampton, Southampton, SO 171 BJ, UK.

# Figures

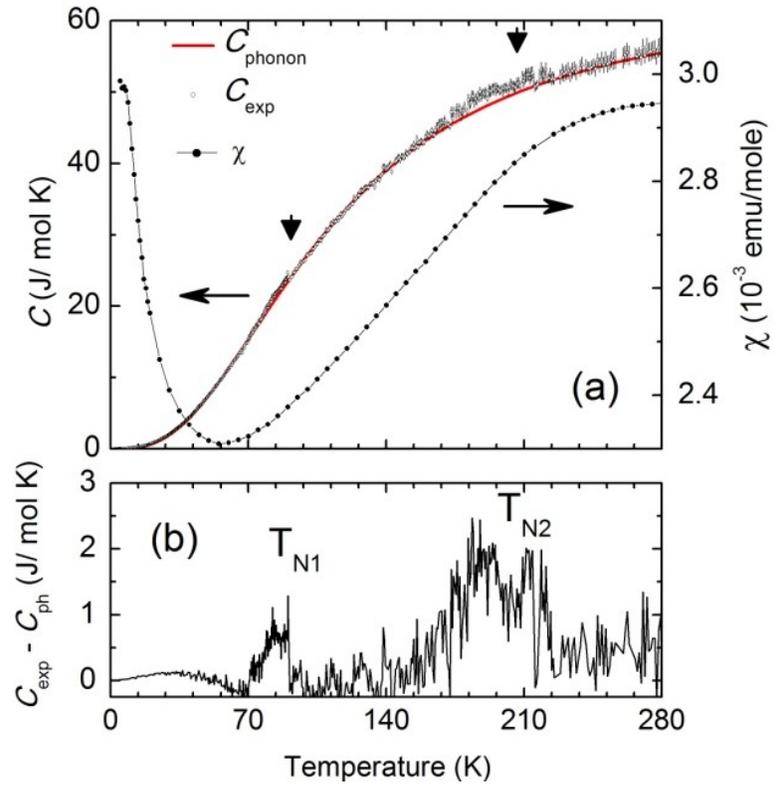

**Figure 1.** Temperature dependent (a) zero-field cooled dc magnetic susceptibility, $\chi(T)$ (right-axis), under an applied field of 20 mT, and the heat capacity, $C(T)$ (left-axis), of β-NaMnO$_2$. The red line over the $C(T)$ data is the calculated phonon contribution to the specific heat, $C_{ph}(T)$ (see text). (b) The heat capacity remaining after subtracting the $C_{ph}(T)$ contribution from the experimental $C(T)$ depicts two anomalies assigned as $T_{N1}$ and $T_{N2}$.

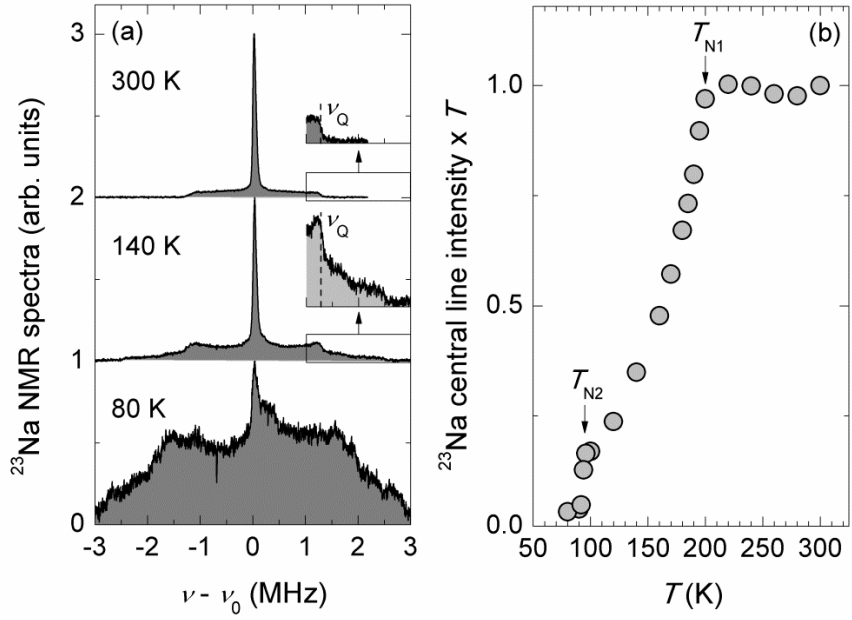

**Figure 2.** (a) Normalized $^{23}$Na NMR powder spectra of β-NaMnO$_2$ revealing two different magnetic regimes that evolve with temperature lowering. The spectra are shifted vertically for clarity. The insets point to a specific part of the spectra, where the quadrupolar frequency is indicated by the vertical dashed line. (b) The temperature dependence of the $^{23}$Na NMR central line intensity multiplied by temperature for β-NaMnO$_2$. The arrows indicate the two transition temperatures $T_{N1}$ and $T_{N2}$.

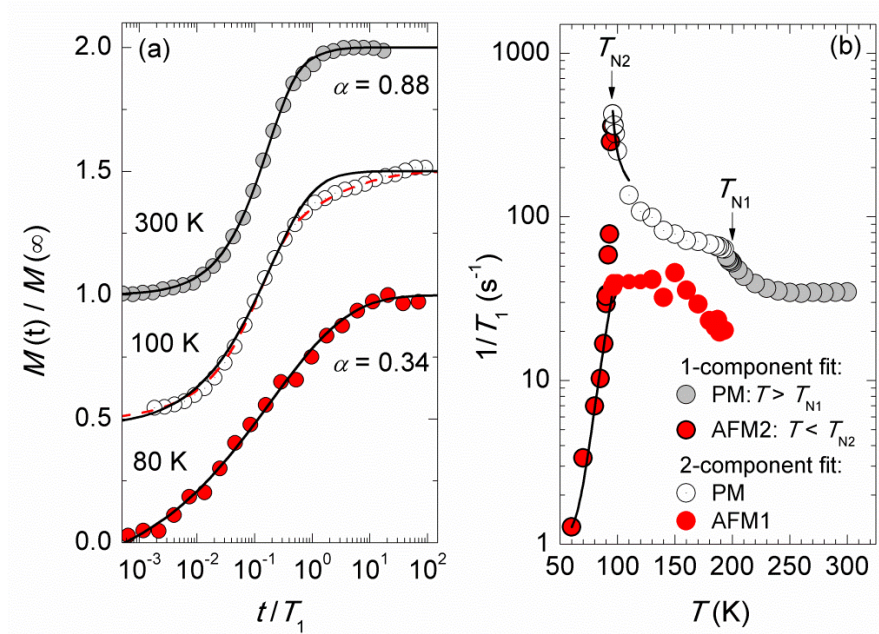

**Figure 3.** (a) Normalized magnetization-recovery curves at a few selected temperatures. The datasets are shifted vertically for clarity. The solid lines are fits of a stretched single-component magnetic-relaxation model for $I= 3/2$ (Eq. 2; see text), while the dashed line corresponds to the fit with two such components. Please note that significantly different stretching exponent α is found for temperatures above $T_{N1}$ and below $T_{N2}$. (b) The temperature dependence of the spin-lattice relaxation rate for β-NaMnO$_2$. The arrows indicate the two transition temperatures. A double-component fit is needed in the intermediate temperature regime $T_{N2} < T < T_{N1}$. The solid lines indicate a critical type of behaviour for $T > T_{N2}$ and an activated one for $T < T_{N2}$ (see text for details).

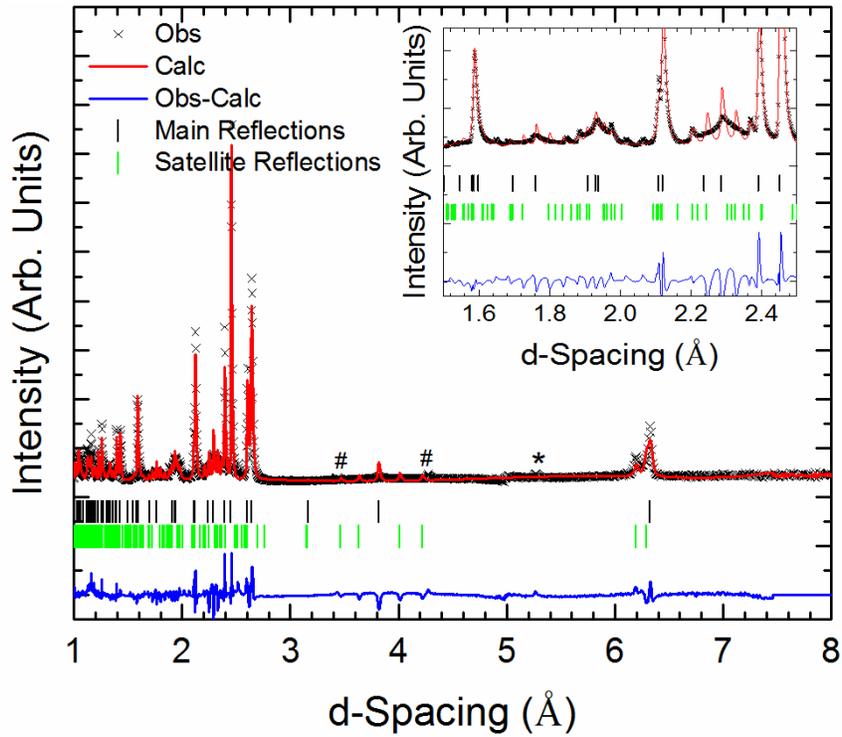

**Figure 4.** Rietveld plot at 300 K for the β-NaMnO$_2$ structure in the *Pmmn(α00)000* superspace group. Inset: zoom of the low *d*-spacing region inferring that stacking faults and defects give rise to a peculiarly broadened profile function. In both panels observed (black crosses), calculated (red line) and difference (blue line) pattern are shown. The tick marks indicate the calculated position of the main (black ticks) and satellite reflections (green ticks). The asterisk marks the main reflection form the α-NaMnO$_2$ impurity, whereas the hash-tags indicate, for example, two satellite peaks that are slightly off with respect to the calculated Bragg position indicating the possibility of the other two components of the modulation vector to be different from zero (see text for details).

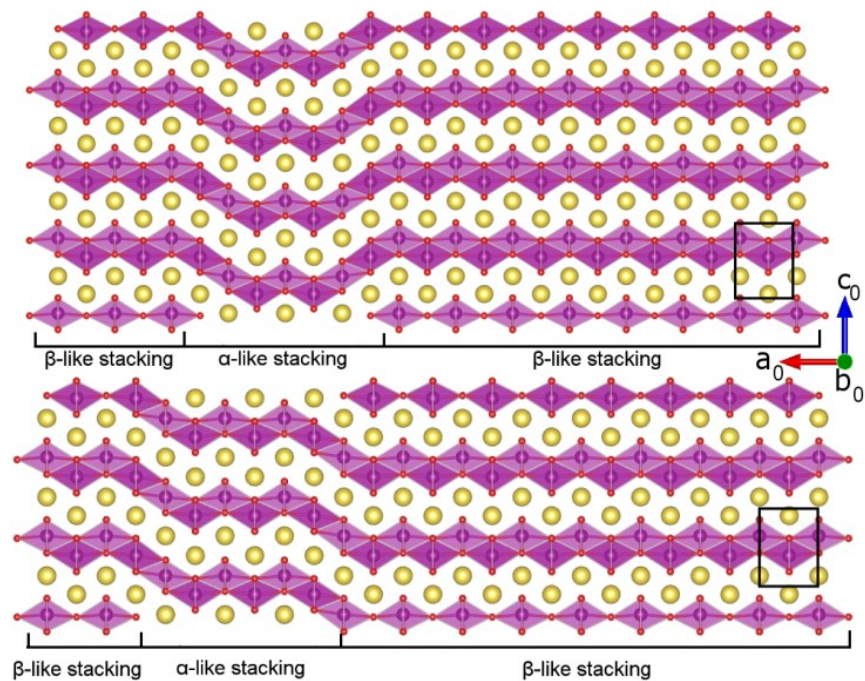

**Figure 5.** Projection of the structure in the ac-plane, depicting the refined incommensurate compositional modulated structure; two types of stacking changing between the $NaMnO_2$ polymorphs are shown. The violet atoms represent the Mn, the yellow ones the Na, and the red spheres the oxygen atoms. The small rectangle indicates the unit cell of the average *Pmmn* structure (see Figure S1).

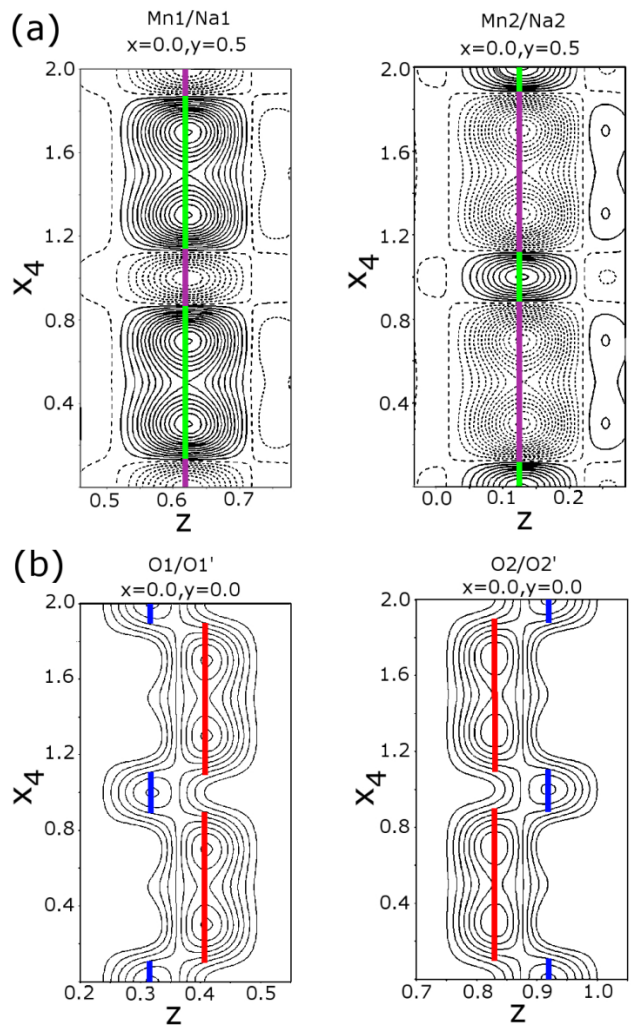

**Figure 6.** Fourier maps of the observed structure factor ($F_{obs}$) depicting the crystallographic cation sites (a) and oxygen positions (b). The solid colored lines represent the calculated position of the atoms showing no positional modulation along the $x_4$ for the Mn/Na but its presence for the oxygen sizes (violet Mn, green Na, red oxygen and blue the primed oxygen position). The black continuous lines indicate the positive density iso-surface and the dashed lines the negative iso-surface (the neutron scattering length for the Mn atoms is negative). The iso-surface contours correspond to 2 scattering density units (Å$^{-2}$) in all the plots.

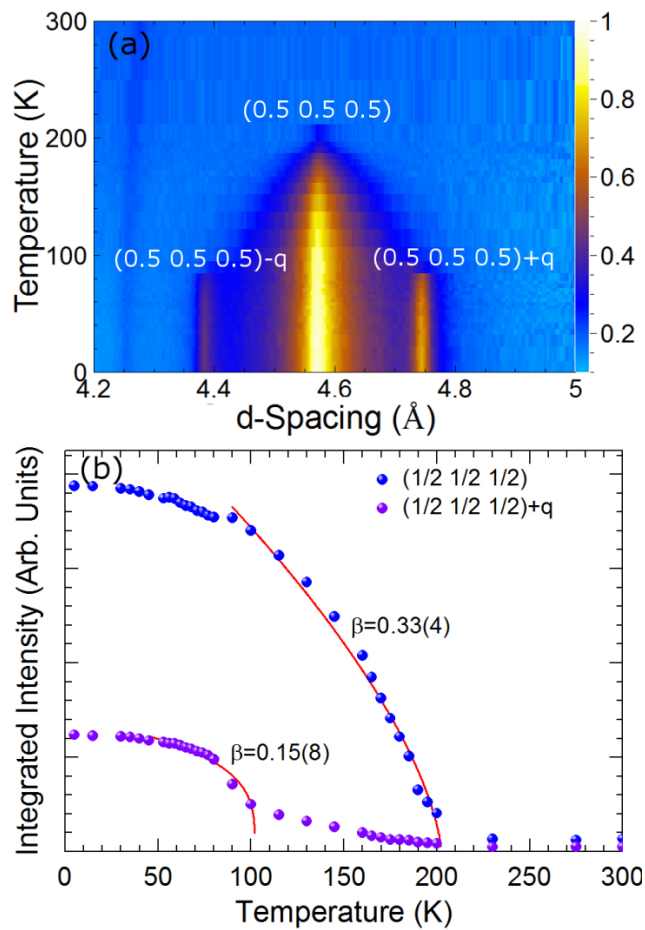

**Figure 7.** (a) A long d-spacing section of the neutron powder diffraction patterns as a function of temperature, showing the complex nature of the magnetic contribution to the pattern. Color map: the neutron scattering intensity. (b) Integrated intensity versus temperature for the main magnetic reflections with propagation vector **k**= (½, ½, ½), and for the satellites with propagation vector **k+q**, where **q** = (0.077(1),0,0). The lines over the data points depict the fit to the critical region (see text).

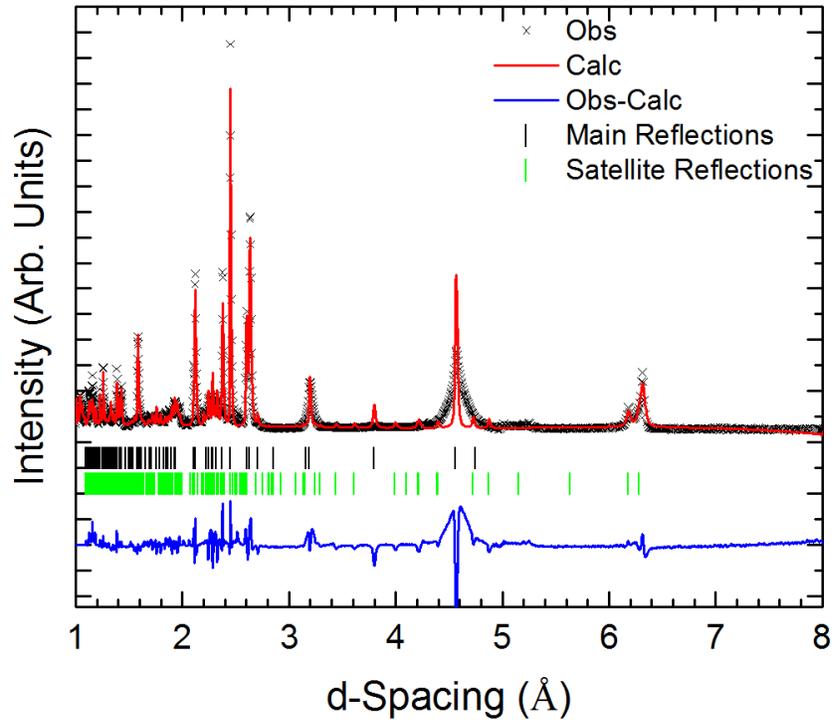

**Figure 8.** Rietveld plot at 100 K of the β-NaMnO$_2$ structure in $C_a2'/c'(\alpha 0\gamma)00$ superspace group, with cell parameters a= 5.7108(2) Å, b= 12.6394(9) Å, c= 5.5397(4) Å, β= 120.96(7)°, and **q**= (0,0, 0.078(1)). Observed (black crosses), calculated (red line) and difference (blue line) patterns are reported. The tick marks indicate the calculated position of the main (black ticks) and satellite reflections (green ticks).

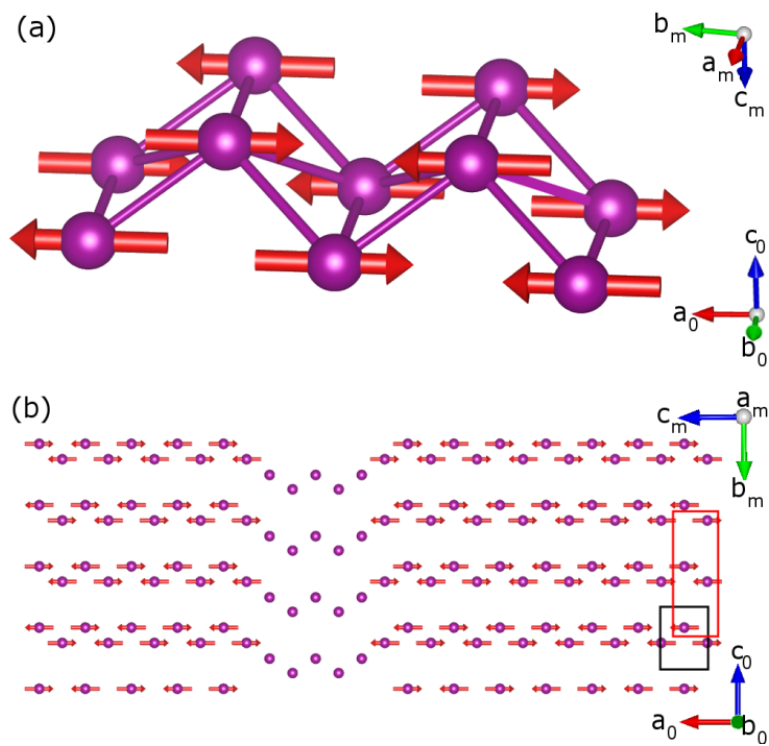

**Figure 9.** Sketch of the magnetic structure below 200 K, (a) along the Mn zig-zag chain typical of the β- polymorph ($\mathbf{a_o}$ direction) and (b) in the same projection as for Figure 6 (top panel). The black rectangle depicts the unit cell of the average *Pmmn* structure ($a_0$= 4.7851(2) Å, $b_0$= 2.85699(8) Å, $c_0$= 6.3287(4) Å), while the red rectangle indicates the unit cell of the average low temperature monoclinic structure ($a_m$= 5.7112(2) Å, $b_m$= 12.6388(9) Å, $c_m$= 5.5365(4) Å, β= 120.97(7)°); please note that the $\mathbf{c_m}$-axis is inclined by ~60° out of the plane.

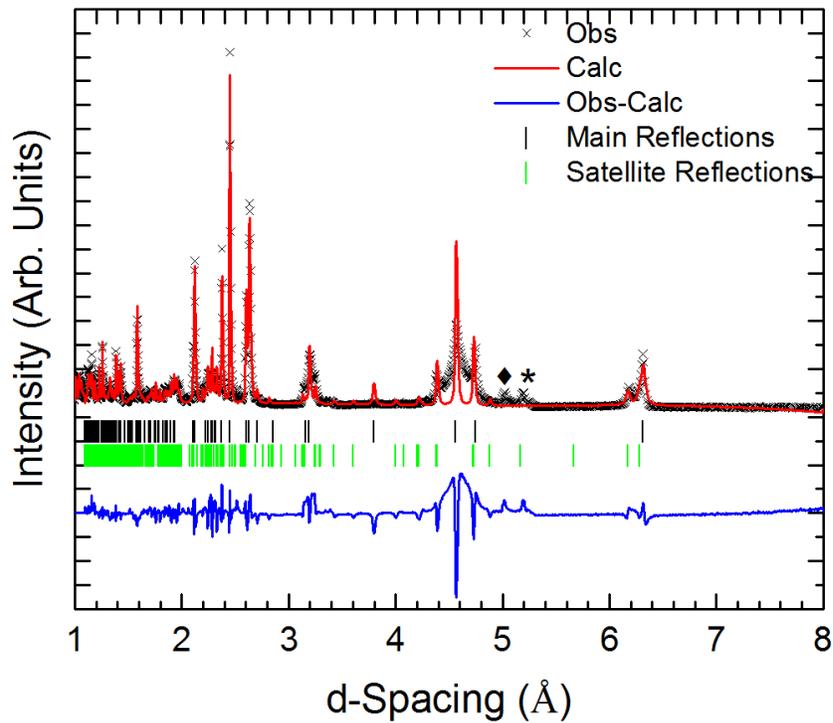

**Figure 10.** Rietveld plot at 5 K for the β-NaMnO$_2$ structure in $C_a2'/c'(\alpha 0\gamma)00$ superspace group, with cell parameters a= 5.7112(2) Å, b= 12.6388(9) Å, c= 5.5365(4) Å, β= 120.97(7)°, and ***q***=(0, 0, 0.081(1)). Observed (black crosses), calculated (red line) and difference (blue line) patterns are shown. The tick marks indicate the calculated position of the main (black ticks) and satellite reflections (green ticks). The asterisk marks the main nuclear and magnetic reflections from the α-NaMnO$_2$ impurity phase, whereas the diamond indicates the main MnO magnetic reflection.

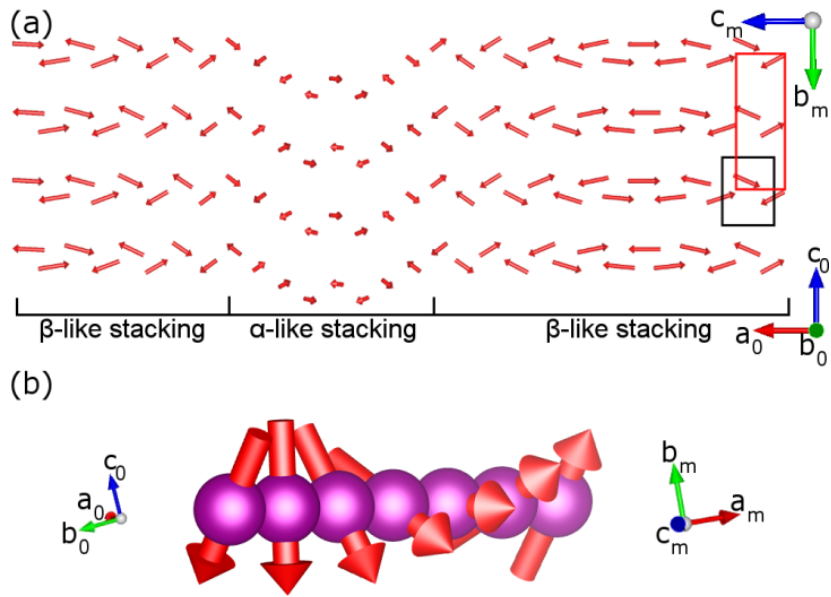

**Figure 11.** (a) Schematic of the β-NaMnO$_2$ modulated magnetic structure at 5 K, projected at the same plane as the nuclear structure shown in Figure 6 (top panel). (b) Sketch of the incommensurate part of the magnetic structure depicting a proper screw order propagating along the (-110) direction with respect to the average *Pmmn* unit cell. In both panels the axes directions with subscript '0' indicate the average orthorhombic *Pmmn* cell (black rectangle), whereas the axes with subscript 'm' indicate the direction of the low-temperature monoclinic structure (red rectangle).

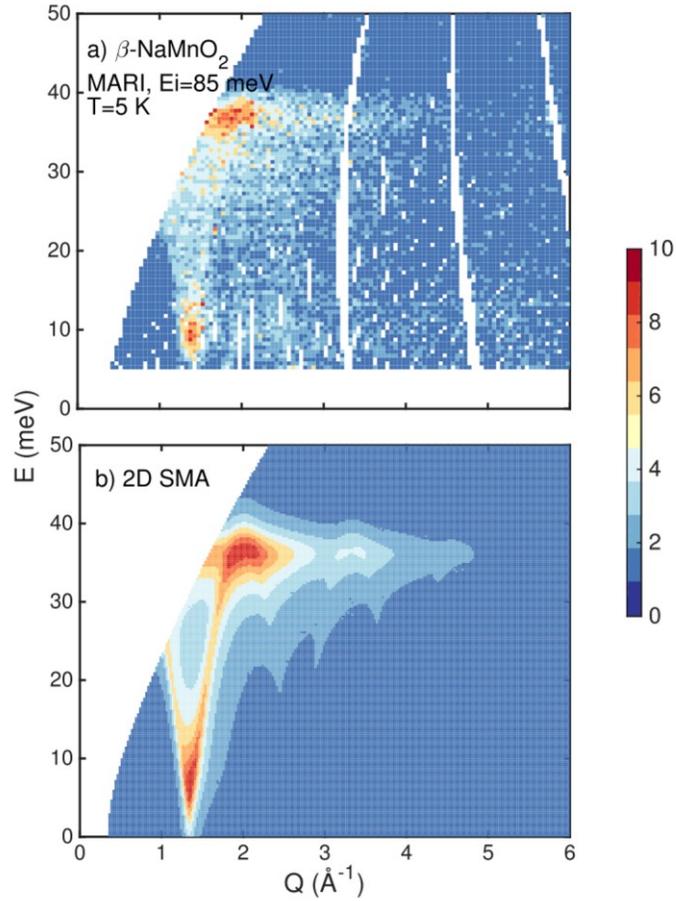

**Figure 12.** (a) The powdered averaged magnetic scattering in β-NaMnO$_2$ and (b) the corresponding Single Mode Approximation (SMA) heuristic model, with two-dimensional (2D) interactions. The background subtraction method to remove phonon scattering and instrument background are described in the text. Color map: indicates the powder average scattering intensity $\bar{I}(\vec{Q}, \hbar\omega)$ (see text for details).

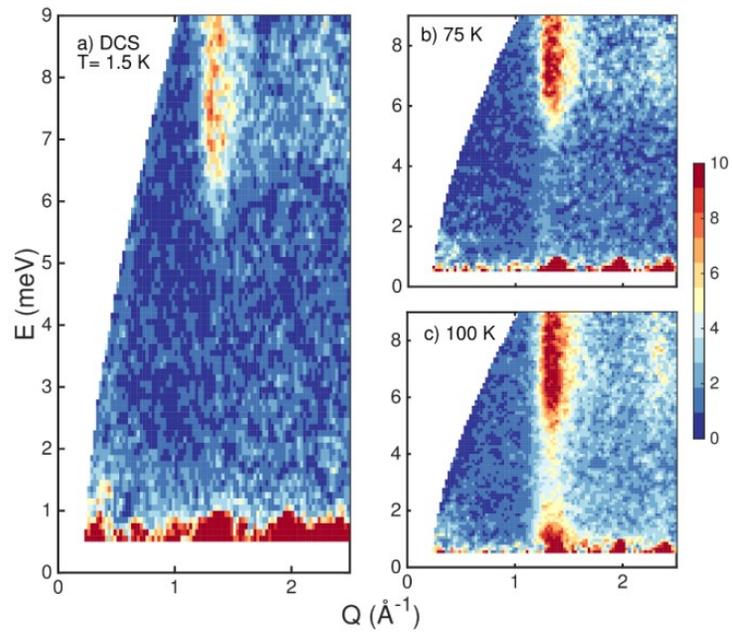

**Figure 13.** The temperature dependence of the low-energy magnetic fluctuations in β-NaMnO$_2$, measured on the high-resolution DCS spectrometer. All data has been corrected for a temperature independent background using the detailed balance relation. Color map: indicates the powder average scattering intensity.

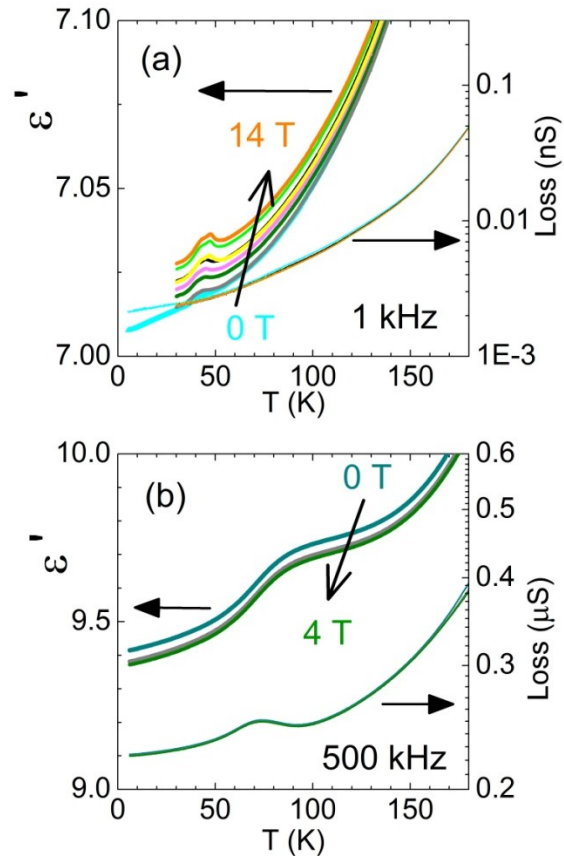

**Figure 14.** Temperature dependent dielectric permittivity, $\varepsilon'(T)$, as a function of the applied magnetic field for α-NaMnO$_2$ (a) and β-NaMnO$_2$ (b).